\documentclass[aps,pra,twocolumn,superscriptaddress]{revtex4}

\usepackage{graphicx}
\usepackage{amsmath}
\usepackage{braket}
\usepackage{color,epsfig}
\usepackage{amssymb}
\usepackage{multirow}
\begin{document}


\title{Thermodynamics and magnetism in the 2D-3D crossover of the Hubbard model}


\author{Eduardo Ibarra-Garc\'ia-Padilla}
\email[]{eibarragp@rice.edu}
\affiliation{Rice Center for Quantum Materials, Department of Physics, Rice University, Houston, TX 77005, USA.}
\author{Rick Mukherjee}
\affiliation{Department of Physics, Imperial College London, SW7 2AZ London, United Kingdom}
\author{Randall G. Hulet}
\affiliation{Rice Center for Quantum Materials, Department of Physics, Rice University, Houston, TX 77005, USA.}
\author{Kaden R. A. Hazzard}
\affiliation{Rice Center for Quantum Materials, Department of Physics, Rice University, Houston, TX 77005, USA.}
\author{Thereza Paiva}
\affiliation{Departamento de F\'isica dos S\'olidos, Instituto de
F\'isica, Universidade Federal do Rio de Janeiro, 21945-970, Rio de
Janeiro, RJ, Brazil}
\author{Richard T. Scalettar}
\affiliation{Department of Physics, University of California, Davis, CA 95616, USA.}



\date{\today}

\begin{abstract}
The realization of antiferromagnetic (AF) correlations in ultracold fermionic atoms on an optical lattice is a significant achievement. Experiments have been carried out in one, two, and three dimensions, and have also studied anisotropic configurations with stronger tunneling in some lattice directions. Such anisotropy is relevant to the physics of cuprate superconductors and other strongly correlated materials. Moreover, this anisotropy might be harnessed to enhance AF order.  Here we numerically investigate, using Determinant Quantum Monte Carlo, a simple realization of anisotropy in the 3D Hubbard model in which the tunneling between planes, $t_\perp$, is unequal to the intraplane tunneling $t$. This model interpolates between the three-dimensional isotropic ($t_\perp = t$) and two-dimensional ($t_\perp =0$) systems. We show that at fixed interaction strength to tunneling ratio ($U/t$), anisotropy can enhance the magnetic structure factor relative to both 2D and 3D results. However, this enhancement occurs at interaction strengths below those for which the N{\'e}el temperature $T_{\rm N\acute{e}el}$ is largest, in such a way that the structure factor cannot be made to exceed its value in isotropic 3D systems at the optimal $U/t$. We characterize the 2D-3D crossover in terms of the magnetic structure factor, real space spin correlations, number of doubly-occupied sites, and thermodynamic observables. An interesting implication of our results stems from the entropy's dependence on anisotropy. As the system evolves from 3D to 2D, the entropy at a fixed temperature increases. Correspondingly, at fixed entropy, the temperature will decrease going from 3D to 2D. This suggests a cooling protocol in which the dimensionality is adiabatically changed from 3D to 2D.
\end{abstract}

\maketitle

\section{Introduction}

Quantum simulation uses engineered quantum systems, such as ultracold atoms in lattices, to realize many-body models of interest in ways that offer powerful control over the system and probes of its physics \cite{Bloch2012,Gross2017,Altman2019}. A prototypical example is using fermions in an optical lattice as an optical lattice emulator (OLE) to realize the Fermi-Hubbard model \cite{Joerdens2008,Schneider2008,Strohmaier2010,
Esslinger2010,Hart2015,Cocchi2016,Cocchi2017,Tarruell2018}. Such simulations allow experiments to flexibly tune the kinetic and interaction energies, lattice geometry, and lattice filling, and in principle use this control to study antiferromagnetism (AF), superconductivity, pseudogap, and strange metal behavior, for example.

AF is intriguing in its own right and is a natural first step to more exotic phases \cite{Chiu2019,Brown2019,Huang2019}. AF in cold atoms has been studied in bosonic atoms \cite{Simon2011}, spin-1/2 ions \cite{Kim2010,Britton2012}, in highly anisotropic lattices \cite{Greif2013,Imriska2014,Greif2015,Ozawa2018}, and in other more recent theoretical work \cite{Kromer2014, Lenz2016,Raczkowski2012,Kung2017,Ehlers2018,Paiva2010}. 
In a fermion OLE, spin-selective Bragg scattering observed AF correlations at temperatures down to $1.4 \, T_{\rm N\acute{e}el}$ in a three-dimensional (3D) cubic lattice \cite{Hart2015}, where $T_{\rm N\acute{e}el}$ is the N{\'e}el temperature (the critical temperature for AF ordering), with an accompanying characterization of the Mott insulator equations of state \cite{Duarte2015}. In addition, quantum gas microscopy \cite{Bakr2009,Sherson2010,Weitenberg2011,Endres2011,Endres2013,Greif2016,Yamamoto2016,Okuno2020} has provided direct observation of correlations beyond nearest-neighbors, through real-space imaging of AF order in one \cite{Boll2016} and two \cite{Parsons2016,Cheuk2016,Mazurenko2017} dimensions. As we will elaborate on later, dimensionality plays an important role in the transition temperature to the antiferromagnetic phase, being equal to zero in 2D, but finite in 3D.

Although OLEs are giving us new insights into quantum matter, there are also significant challenges. Of particular relevance here is that, although experiments have achieved spin correlations which extend across the finite 2D lattice \cite{Mazurenko2017}, so far experiments have not reached sufficiently low temperatures or entropies to observe a long-range ordered AF phase in a regime where $T_{\rm N\acute{e}el}>0$, i.e. where correlations would persist to long range as the system size is increased arbitrarily. In order to achieve this goal, several cooling protocols exist. One that has received a lot of attention from both theory and experiment is to use spatial subregions as repositories for excess entropy, allowing for lower temperatures in other regions \cite{Ho2009,Mazurenko2017,Greif2013,Haldar2014,Chiu2018}, but reaching the N{\'e}el temperature, and below, remains an outstanding challenge.

Anisotropic systems that have larger tunneling rates in some directions than others offer potentially richer varieties of physics than simple 1D, 2D, or 3D cubic lattices. Anisotropic systems are relevant to real materials, as discussed below, while also suggesting a route to achieving longer-range AF order. Specifically, it is known that 2D systems offer stronger nearest neighbor correlations for a given entropy than 3D systems \cite{Greif2013,Greif2015}, making them favorable to search for short-ranged AF. However, true long-range order cannot develop at $T>0$ in 2D due to the Mermin-Wagner theorem, in contrast to 3D. Thus a potential scenario for anisotropic lattices that interpolate between 2D and 3D is that they retain the strong AF  correlations associated with 2D planes, while being able to develop long-range order by virtue of the interplane tunnelings.

This paper explores the evolution of AF correlations in the half-filled repulsive Hubbard model across the 2D-3D crossover using Determinant Quantum Monte Carlo (DQMC) \cite{Blankenbecler1981,Sorella1989}. DQMC~\cite{Hart2015,Brown2019,Paiva2010,Duarte2015,Cheuk2016,Mazurenko2017,
Mitra2017,Brown2017,Brown2018,Chan2020} and other numerical solutions of the Hubbard model [such as numerical linked-cluster expansion (NLCE)~\cite{Duarte2015,Cheuk2016,Brown2017}, dynamic mean-field theory (DMFT)~\cite{Dare2007,DeLeo2011,Joerdens2010,Brown2018}, density matrix renormalization group (DMRG)~\cite{Manmana2011}, and diagrammatic QMC~\cite{Burovski2006,Kozik2013}] have provided key input in the interpretation of experiments and, in particular, in the determination of temperature. In this paper, the evolution of AF correlations is characterized as a function of both temperature $T$ and entropy $S$, to allow for a deeper understanding of the optimization of AF at fixed $S$.

An important conclusion is that, for interaction strength $U$ less than (roughly) the 2D bandwidth, the long-range AF correlations at a given temperature or entropy, measured by the magnetic structure factor at the $\vec{k} = (\pi,\pi,\pi)$ wavevector, are maximized in lattices which straddle dimensionality.  Although anisotropy can increase the structure factor at small $U$, it never exceeds the value in the isotropic 3D system evaluated at the optimal $U$. Similar conclusions were reached in Ref.~\cite{Imriska2014} for the 1D-3D crossover using a dynamical cluster approximation (DCA).

In addition to the possibility of achieving AF in OLE, an understanding of dimensional crossover is relevant to strongly correlated materials \cite{Klebel2020}.  Perhaps the most important example is the cuprate superconductors, layered materials for which the superexchange coupling ${J_\perp = 4 t_\perp^2/U}$ between planes is several orders of magnitude lower than the in-plane superexchange ${J= 4 t^2/U}$ \cite{Chakravarty1988,Chubukov1994}. Despite this large anisotropy, $J_\perp$ is crucial to the physics, since in a purely 2D geometry $T_{\rm N\acute{e}el}=0$.

The remainder of this paper is organized as follows: Section~\ref{sec:Hubbard_DQMC} presents the Hubbard Hamiltonian and defines the observables we consider. Section~\ref{sec:Results} presents the main results. Section~\ref{sec:Conclusions} concludes.

\section{The Hubbard Hamiltonian and DQMC}\label{sec:Hubbard_DQMC}

\begin{figure}[htbp!]
	\includegraphics[width=8cm]{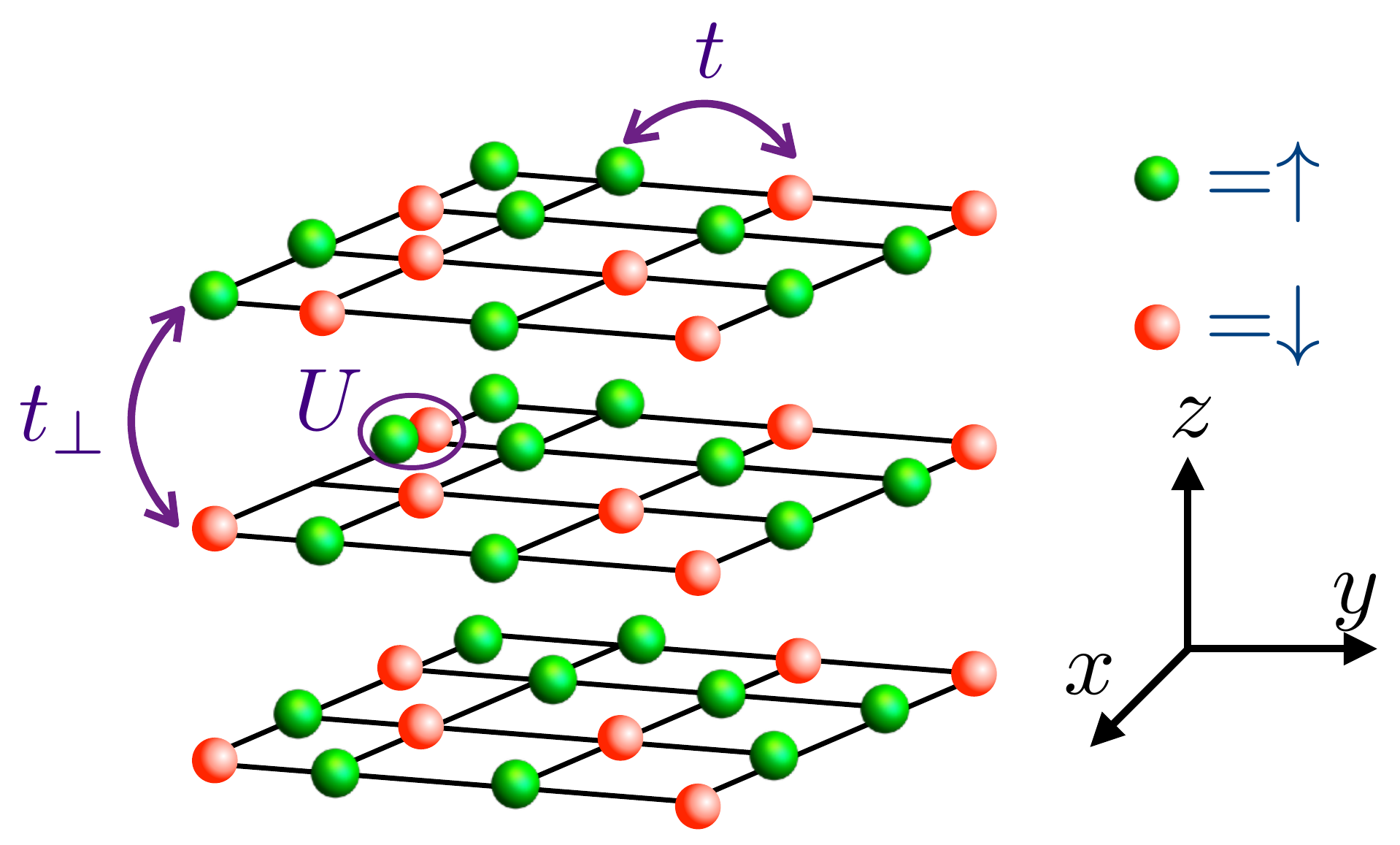}
	\caption{Setup: Fermionic atoms trapped in a three-dimensional anisotropic lattice. Each atom can either be in $\ket{\uparrow}$ state (green) or $\ket{\downarrow}$ state (red) realized using hyperfine states of ultracold atoms. An atom hops to its neighboring site at rate $t$ within a plane and at rate $t_{\perp}$ between the planes. Two atoms with opposite spin states can occupy the same site with energy $U$. The crossover from a three-dimensional lattice to a two-dimensional lattice is achieved by reducing $t_{\perp}$.\label{fig:setup}}
\end{figure}

We investigate the half-filled, anisotropic Hubbard Hamiltonian (depicted in Fig. \ref{fig:setup}),
\begin{align}
\mathcal{H} =  &-t \sum_{\langle ij \rangle_\parallel, \sigma} 
\left( c^{\dagger}_{i\sigma} c^{\phantom{\dagger}}_{j\sigma}
 + \mathrm{h.c.} \right) 
\nonumber
-t_\perp 
\sum_{\langle ij \rangle_\perp,\sigma} \left( c^{\dagger}_{i\sigma}
c^{\phantom{\dagger}}_{j\sigma}
     + \mathrm{h.c.} \right) \nonumber
\\
&+U \sum_{i} \left(n_{i\uparrow}-\frac{1}{2}\right) \left(n_{i \downarrow}-\frac{1}{2}\right),
\label{eq:ham}
\end{align}
in which hopping $t$ connects pairs of sites $\langle ij \rangle_\parallel$ which are neighbors in the same plane of a 3D cubic lattice, while a weaker hopping $t_\perp < t$ connects pairs of sites $\langle ij \rangle_\perp$ which are neighbors in adjacent planes.  $U$ is the on-site repulsion between fermions of opposite spin.  The limits $t_\perp=t$ and $t_\perp=0$ correspond to the 3D and 2D Hubbard Hamiltonians respectively. The chemical potential is set to ${\mu=0}$. This choice of $\mu$ in Eq.~\eqref{eq:ham} gives half-filling on average, i.e. ${\braket{n_i}= \braket{n_{i\uparrow}} + \braket{n_{i\uparrow}} = 1}$ for all values of $t, \, t_\perp, \, U$, and temperature $T$. At half-filling, DQMC is free of the sign-problem, and as consequence, low temperature physics can be accessed. We set $k_B=1$ throughout. 

\begin{figure*}[tbp!]
	\includegraphics[width=\linewidth]{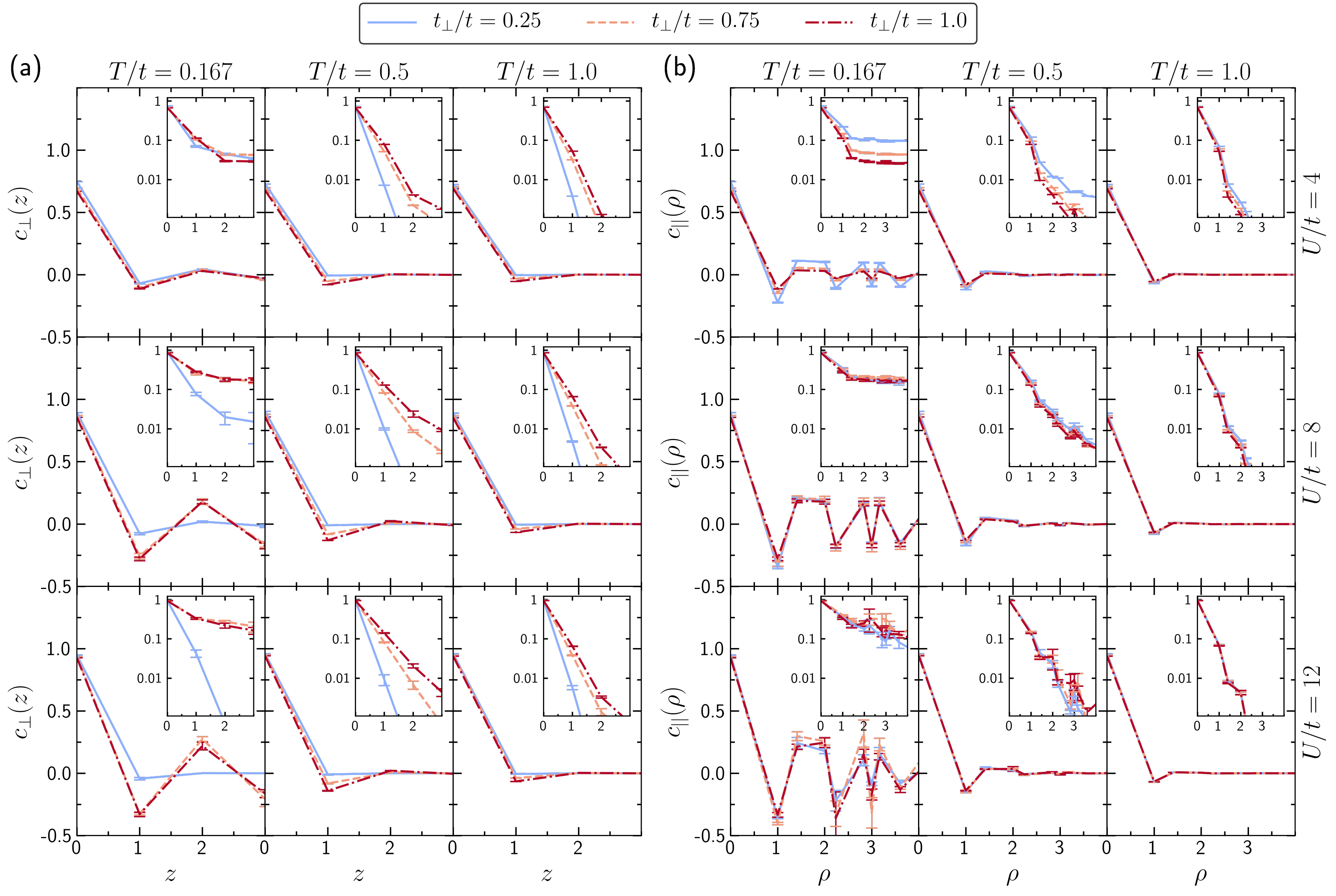}
	\caption{(a) Out-of-plane $c_{\perp}(z)$ and (b) in-plane $c_{||}(\rho)$ spin correlations as a function of separation for different values of $U/t$ and $t_{\perp}/t$. Insets show the same plot with the correlations on a log scale. \label{fig:corr_spatial}}
\end{figure*}

We are interested both in the thermodynamics, e.g. energy and entropy,  how the temperature $T$ changes with $t_\perp$ at fixed entropy $S$, and also with the behavior of the real space spin-spin correlation function ${c(\vec r \,)=  c(\rho,z)}$ (where $\rho$ is the magnitude of the in-plane components of $\vec{r}$), in particular the in-plane $c_{||}(\rho)$ and out-of-plane $c_{\perp}(z)$ correlation functions, as well as the magnetic structure factor $S(\vec q \,)$:
\begin{align}
c(\vec r \,)  &=   \big\langle  \,
\big(n_{\vec r_0+\vec r,\uparrow} - n_{\vec r_0+\vec r,\downarrow}\big) 
\,\big(n_{\vec r_0,\uparrow} - n_{\vec r_0,\downarrow}\big) 
\, \big\rangle,
\nonumber \\
c_{||}(\rho) &= c(\rho,z=0), \nonumber \\
c_{\perp}(z) &= c(\rho=0,z), \nonumber \\
S(\vec q \,) &= \sum_{\vec r} e^{i \vec q \cdot \vec r} \, c(\vec r \, ),
\label{eq:crandsq}
\end{align}
where these averages are taken in thermal equilibrium at fixed temperature $T$ and chemical potential ${\mu=0}$. The structure factors can diagnose long range order. At half-filling, the Fermi surface is nested for any $t_\perp$, so that the ordering wavevector is always $(\pi,\pi,\pi)$ regardless of the degree of anisotropy. For that reason we focus on the AF structure factor $S\big(\vec q=(\pi,\pi,\pi)\big)$,  which we denote $S_\pi$. In addition, Ref.~\cite{Xu2013} contains a mean field theory study of the crossover from 3D to 2D considered here, including careful treatment of finite-size and shell effects to ensure the correct ordering wavevector is captured at all densities.

The averages of thermal equilibrium observables of Eq.~\eqref{eq:ham} are evaluated with DQMC \cite{White1989} in ${10 \times 10}$ (for ${t_\perp =0}$) and ${6 \times 6 \times 6}$ (for ${t_\perp  > 0}$) lattices. In this method, the introduction of a space- and imaginary time-dependent auxiliary field allows tracing over the fermion degrees of freedom analytically. The auxiliary field is then sampled stochastically. To achieve accurate results, we obtain DQMC data for 20-50 different random seeds for $T/t \leq 1$ and for 1-10 different random seeds for $T/t > 1$. In each realization, 500 sweeps updating the auxiliary field at every lattice site and imaginary time are performed for equilibration and 5000 sweeps for measurements. For each Monte Carlo trajectory measurements of the $\langle S^zS^z\rangle$ and  $\langle S^xS^x \rangle$ correlation functions are made. These are equal on average by the SU(2) symmetry, and both are included in the statistics. The inverse temperature interval $(0,\beta)$ is discretized in steps of $\Delta \tau$ with a Trotter step ${\Delta \tau = 0.05/t}$ for ${U/t=4,8}$ and ${\Delta \tau = 0.04/t}$ for ${U/t=12}$. The number of  global moves per sweep, which update all the imaginary time slices at a given lattice site, to mitigate possible ergodicity issues \cite{Scalettar1991}, is set to 2 for ${U/t=4,8}$ and to 4 for ${U/t=12}$.

Estimates of other systematic errors -- Trotter and finite-size error -- show that the predominant error is statistical, arising from the finite number of measurements. In the following section, error bars are reported as the standard error of the mean for all results. For ${U/t=12}$, where the inverse temperature discretization error is expected to be worst, we can gain insight into the magnitude of this error by considering the difference of the results obtained with Trotter steps  ${\Delta \tau = 0.04/t}$ and  ${\Delta \tau = 0.05/t}$. This difference is below 2.5\% for all observables of interest, comparable to the statistical error in many cases. This discretization error is even smaller for the other two values of $U/t$ considered. Finite-size errors for thermodynamic quantities and nearest-neighbor correlations are estimated by taking the difference between results obtained in cubic lattices with sides of length ${L=4}$ and ${L=6}$ in 3D. These differences are ${\lesssim 5}$\%. At high temperatures and away from the optimal anisotropies, i.e. well above the N\'eel temperature, the error in the structure factor is similar, but for ${T\lesssim T_{\text{N{\'e}el}}}$, the structure factor is sensitive to longer-ranged correlations, including those between sites separated by distances comparable to $L$. Here, finite-size effects can be more significant, ${\sim 50\%}$ in our calculations. (Indeed, below $T_{\text{N\'eel}}$, the difference in $S_\pi$ in a finite and infinite system is infinitely large, and a different extrapolation scheme would be necessary to infer the ${L=\infty}$ results.). Results for the structure factor at low temperatures where it has become independent of temperature should therefore be interpreted with some care. However, we expect the conclusions of our paper to remain. 
A detailed study of finite-size effects in the structure factor can be found in Refs.~\cite{Staudt2000,Kozik2013}, where careful finite-size scaling techniques are used to extract the N{\'e}el temperature in 3D. For more discussion of the finite-size effects in the 2D-3D crossover, see the Appendix \ref{App:Appendix_A}.

\section{Results}\label{sec:Results}

This section shows the main results of this paper. We calculate several observables as functions of $T/t$, $U/t$, and $t_\perp/t$: the spatial correlation functions $c_{||}(\rho)$ and $c_{\perp}(z)$, the AF structure factor $S_\pi$, the double occupancy ${\mathcal{D} = \braket{n_{i,\uparrow} n_{i,\downarrow}}}$, the contributions to the specific heat $C(T)$ from the interaction and kinetic energies, and the entropy per site $S/N$ -- where $N$ denotes the number of sites. All of these observables contain important information about the physics and can be measured in experiments with ultracold atoms. The double occupancy is a key measure of the Mottness and insulating nature of the system, and the correlations and structure factor give  information about the magnetic phase diagram. The thermodynamic observables  give information about the ordering of the state -- its spatial coherence (kinetic energy) and to what extent degrees of freedom are capable of fluctuating (the entropy and specific heat). The entropy is usually obtained by ramping from a weakly interacting gas near-adiabatically, and the entropy of the weakly interacting gas can be determined by thermometry. As the temperature is often not directly experimentally accessible in strongly interacting systems, understanding the dependence of observables on $S$ is crucial.

Figure~\ref{fig:corr_spatial} shows the out-of-plane and in-plane spatial correlations for different values of $U/t$, $T/t$ and $t_\perp/t$. First, let's focus on the first row of panels (a) and (b), which corresponds to $U/t=4$. The spatial correlations are larger at small $T/t$, showing clear in- and out-of-plane AF oscillations as a function of distance. At the lowest $T/t$ considered in Fig.~\ref{fig:corr_spatial}, $T/t=0.167$, both $c_{||}(\rho)$ and $c_{\perp}(z)$ indicate a strong antiferromagnetic ordering extending to several lattice sites.  The insets, which present the correlations on a log scale, demonstrate that for the two high temperature sets both $c_{||}(\rho)$ and $c_{\perp}(z)$ have an exponential decay associated with a correlation length $\xi$, while the low temperature data reaches a constant value, an indicator of larger correlation lenght $\xi$ and the onset of long-range order. As one might expect, the strength of the correlations increases as the correlation length increases. All of these trends are similar for the ${U/t=8}$ and ${U/t=12}$ data, but both spatial correlations exhibit stronger AF correlations than for $U/t=4$. 

Now let's focus on how the the low temperature data for panels (a) and (b) evolves with $t_\perp/t$. As $t_\perp/t$ increases, the between-plane correlations get stronger while the in-plane correlations get slightly weaker. The effect on in-plane correlations is strongest for ${U/t=4}$ and nearly negligible for ${U/t=8,12}$.

\begin{figure}[tbp!]
	\includegraphics[width=\linewidth]{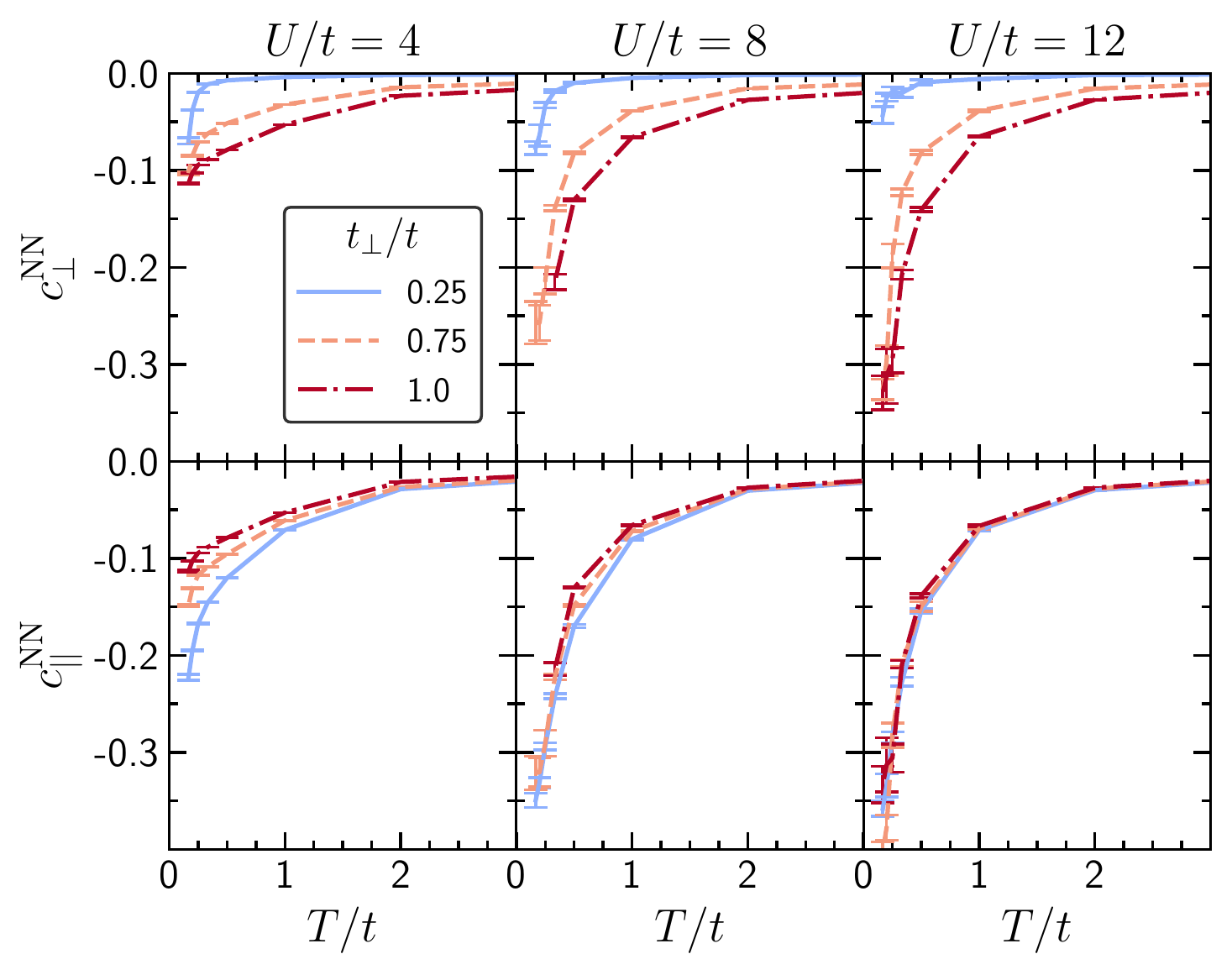}
	\caption{Nearest-neighbor spin correlations as a function of temperature: $c^{\rm NN}_{\perp} = c^{\protect\phantom{\rm N}}_{\perp}(z=1)$ and $c^{\rm NN}_{||} = c^{\protect\phantom{\rm N}}_{||}(\rho=1)$.\label{fig:corrvsT}}
\end{figure}

In Fig. \ref{fig:corrvsT} we plot the in-plane and out-of-plane nearest-neighbor spatial correlations, $c^{\rm NN}_{||}$ and $c^{\rm NN}_{\perp}$, as functions of temperature $T/t$ at various $t_\perp/t$. Both correlation functions get enhanced at small $T/t$ and large $U/t$. Similar to the trends of longer-ranged correlations shown in Fig.~\ref{fig:corr_spatial}, we see that at large $U/t$, the in-plane correlations weakly depend of $t_\perp$, but diminish as $t_\perp$ is increased at weak couplings, while the out-of-plane correlations strongly depend on the anisotropy for all interaction strengths. As expected ${c^{\rm NN}_{\perp} \to 0}$ when ${t_\perp \to 0}$, indicating that the 2D planes are decoupled. 

\begin{figure}[htbp!]
	\includegraphics[width=\linewidth]{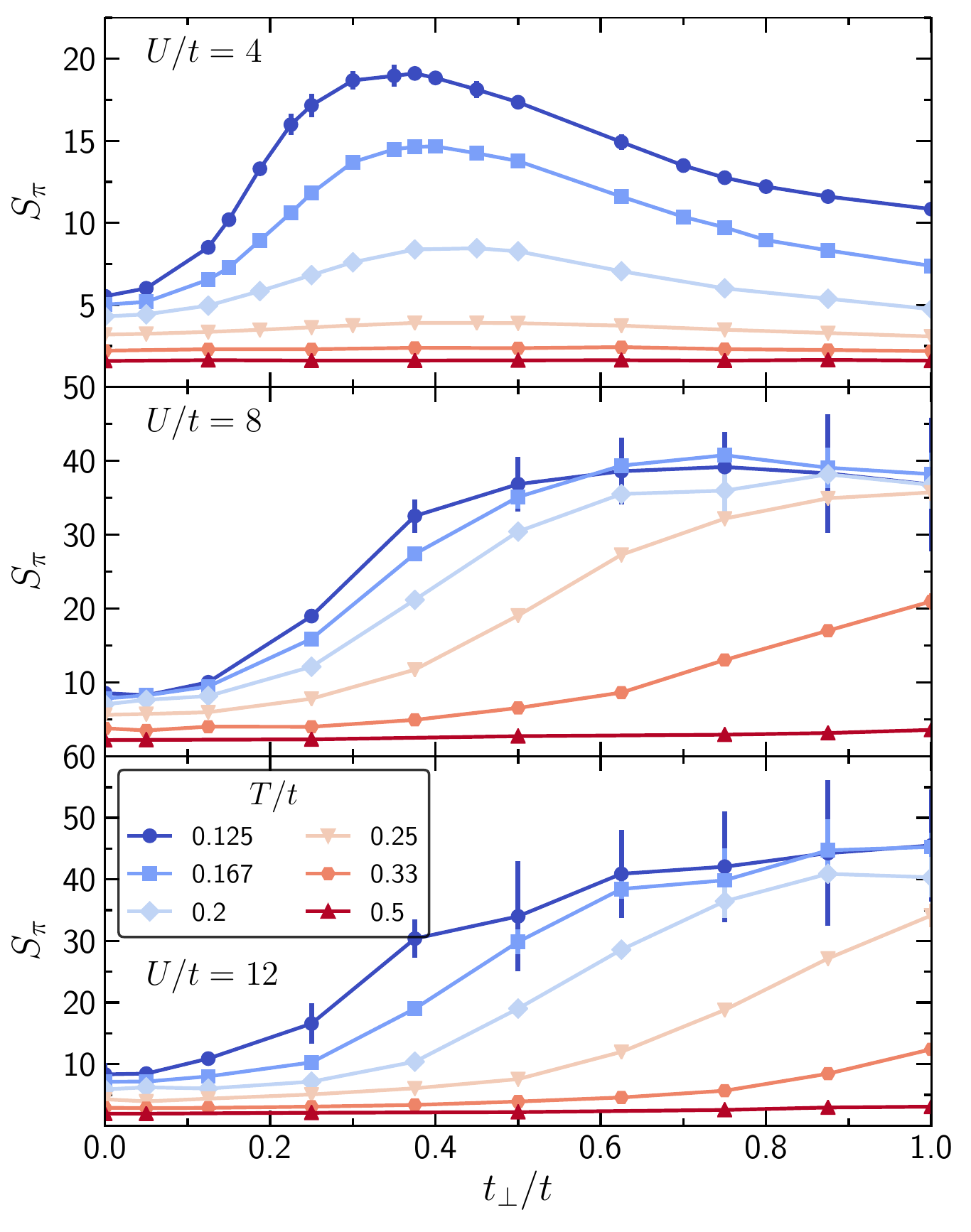}
	\caption{ AF structure factor $S_\pi$ as a function of interplane hopping
		$t_\perp/t$ for different temperatures $T/t$ at $U/t=4$ (top panel),
		$U/t=8$ (middle panel), and $U/t=12$ (bottom panel). At weak coupling, the structure factor increases with anisotropy, i.e. as $t_\perp/t$ decreases from 1.\label{fig:Spivstz}}
\end{figure}

Figure~\ref{fig:Spivstz} presents the structure factor $S_\pi$ vs $t_\perp/t$ at various temperatures. The $U/t=4$ data clearly show that at each temperature, $S_\pi$ is largest between 2D and 3D. In contrast, for $U/t=8$ and $12$, the largest $S_\pi$ occurs at the isotropic point $t_\perp =t$. Although the $U/t=8$ data is consistent with $S_\pi$ being maximized at $t_\perp/t=1$, it is rather independent of $t_\perp/t$ for $t_\perp/t \in (0.5,1.0)$ at the lowest temperatures considered, $T/t \leq 0.167$. Moreover, the maximal $S_\pi$ at $U/t=4$ is smaller than the isotropic $S_\pi$ for $U/t=8$; if one's goal is simply to maximize $S_\pi$ -- irrespective of $U/t$ -- there is no advantage to using anisotropy.

The behavior of $S_\pi$ as a function of $t_\perp/t$ at different interaction strengths, as displayed in Fig.~\ref{fig:Spivstz}, has a simple explanation. In a 3D cubic lattice, $S_\pi$ is maximized around ${U/t \sim 10}$~\cite{Khatami2016}. One effect of anisotropy is to change the average tunneling to be somewhere between $t$ and the smaller $t_\perp$, and thus one would expect anisotropy to decrease the effective tunneling, $t_{\text{eff}}$, and increase the effective $U/t_{\rm eff}$ compared to $U/t$. This change qualitatively explains why $S_\pi$ is maximized around ${t_\perp/t \sim 0.4}$ for $U/t=4$, while is maximized near the isotropic point at $U/t=8,12$.

\begin{figure}[htbp!]
	\includegraphics[width=\linewidth]{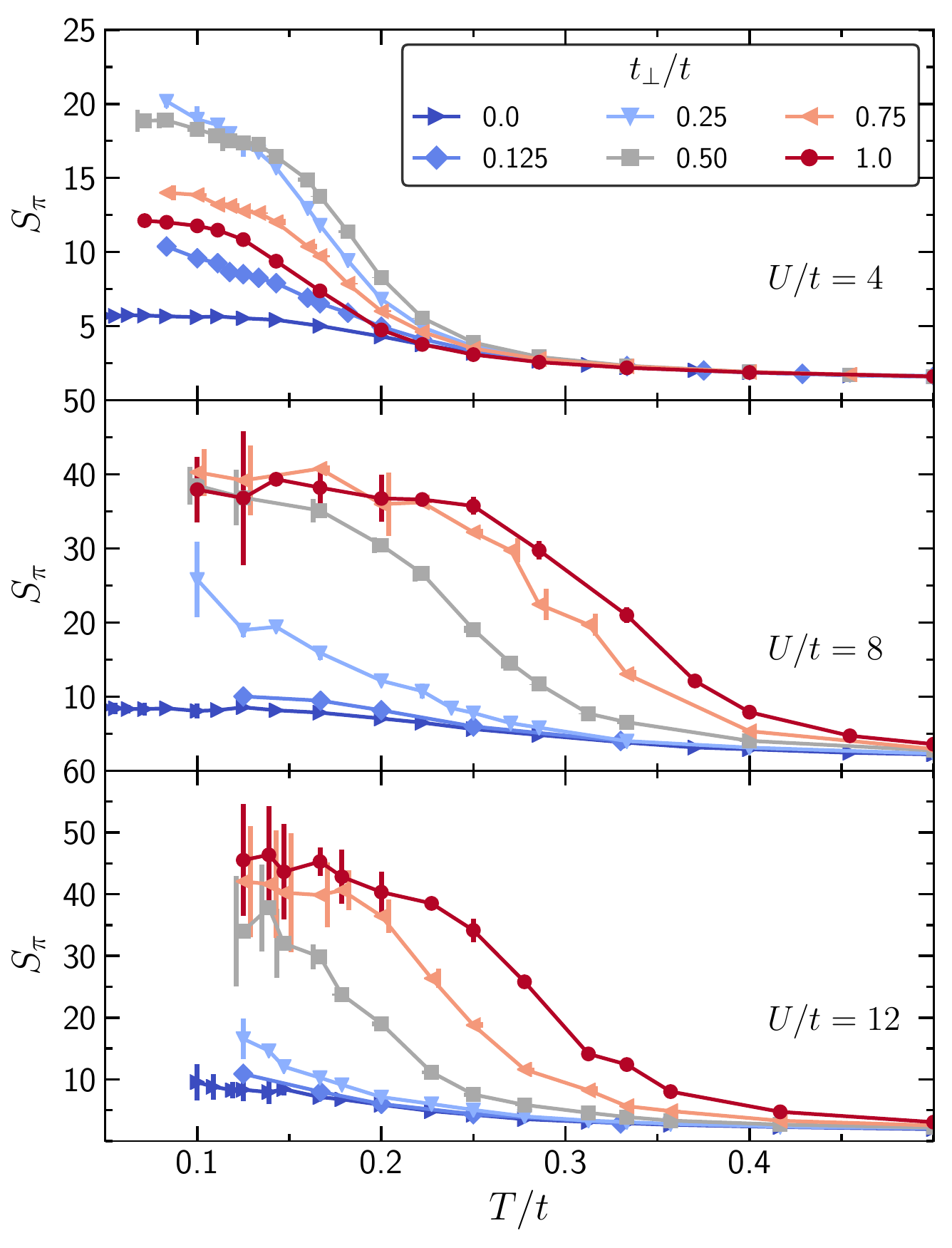}
	\caption{AF structure factor $S_\pi$ as a function of temperature $T/t$ for different interplane hopping $t_\perp/t$.  For $U/t=4$ at low $T/t$, as anisotropy is introduced, $S_\pi$  grows by almost a factor of two down to $t_\perp/t=0.25$. For very strong anisotropy, $t_\perp/t=0.125$, $S_\pi$ comes down and approaches the 2D limit. For $U/t=8,12$, $S_\pi$ decreases with anisotropy. This decrease will overwhelm the benefits of adiabatic cooling (described later; see e.g. Fig. \ref{fig:SvsT1}).
		\label{fig:SpivsT}}
\end{figure}

Figure \ref{fig:SpivsT} shows the AF structure factor $S_\pi$ versus temperature $T/t$ at different $t_\perp/t$. Structure factors at all values of $t_\perp/t$ and $U/t$ grow as temperature is lowered. Generally, the onset of growth of the structure factor begins at the largest temperature for $U/t=8$, although this $U/t$ at which growth onsets can depend on anisotropy. For example, for small ${t_\perp/t}$, ${U/t=4}$ has a similar temperature for the onset of correlations.

\begin{figure}[tbp!]
	\includegraphics[width=\linewidth]{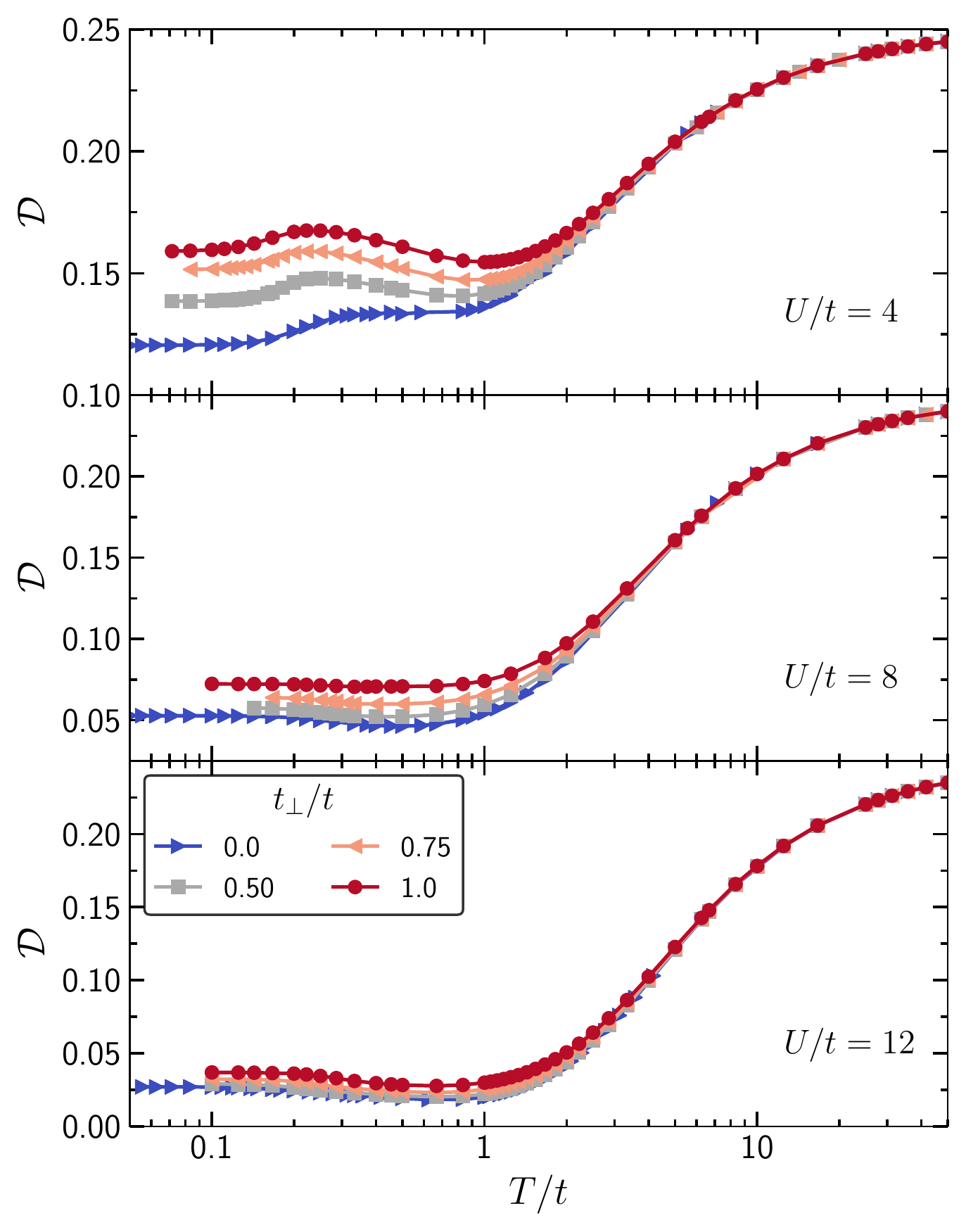}
	\caption{Double occupancy as a function of temperature for $U/t=4,8,12$ at different values of $t_\perp/t$.\label{fig:docc}}
\end{figure}

Figure \ref{fig:docc} plots the double occupancy ${\mathcal{D} = \braket{n_{i,\uparrow} n_{i,\downarrow}}}$ as a function of temperature, and displays three essential parts. Imagine starting at high temperature and cooling the system down. As the temperature is lowered, first the double occupancy $\mathcal D$ goes down. Then, as the temperature is lowered further, it increases (in every case except the 2D ${U/t=4}$). Finally, as the temperature is lowered even further, $\mathcal D$ saturates, or in some cases, such as $U/t=4$, it begins to decrease.

The first feature, the high temperature decrease of $\mathcal D$ upon cooling, is straightforward to understand. At temperatures $T\gtrsim U$, eigenstates with significant numbers of double occupancies will be created, while at temperatures below this, the eigenstates relevant to the $\mu=0$ state will have only a small admixture of doublons, at least for reasonably strong interactions.

The second feature is more interesting, and arises from spin-ordering. We can gain a simple understanding of this starting from the $U\gg t$ limit. For temperatures $T\ll U$, we can think of the states as essentially having a single particle per site with small admixtures of other states. The relevant states in the relevant sector are just determined by their spin configurations. AFM aligned spin configurations will have energy $\propto -t^2/U$ per site lower energy than FM aligned spins. Therefore, as the temperature is lowered below $T\lesssim -t^2/U$, the AFM aligned states become favored. Now consider the doublon content of these two classes of states. The number of doublons in the state with FM aligned spins is small (zero if all the spins are exactly aligned) since Pauli exclusion prevents tunneling. In contrast, there is an admixture $\propto (t/U)^2$ of doublons in the AFM state; it is precisely this admixture which allows some  delocalization of particles that lowers the energy of the AFM states relative to the FM ones. Therefore, as the temperature is lowered, the AFM states are increasingly favored and the number of doublons increases by an amount $\propto (t/U)^2$. (This is why, in general at low temperature, the increase in $\mathcal D$ is accompanied by a lowering of the kinetic energy.) We note that a simple place to check this argument is in a two-site system, where the calculation can be done analytically.

These arguments provide an understanding of the decrease in $\mathcal D$ as $T$ is lowered below $U$ and its small increase (in almost all cases) when $T\lesssim t^2/U$.  This also explains some of the dependences on parameters. For example, the low-temperature value of $\mathcal D$ decreases as $U/t$ increases, and  increases with $t_\perp$. However, some features remain unexplained: Why does $\mathcal D$ the $U/t=4$ decrease again with decreasing temperature at sufficiently low temperatures? And why is there no (visible) increase in $\mathcal D$ with decreasing temperature for the one set of parameter values (${U/t=4}$ for ${t_\perp=0}$).   A simple theory  capturing these more refined features and dependences could  provide  powerful insights into the Hubbard model's physics, and our data will be an excellent test for any candidate theories.

\begin{figure}[tbp!]
	\includegraphics[width=\linewidth]{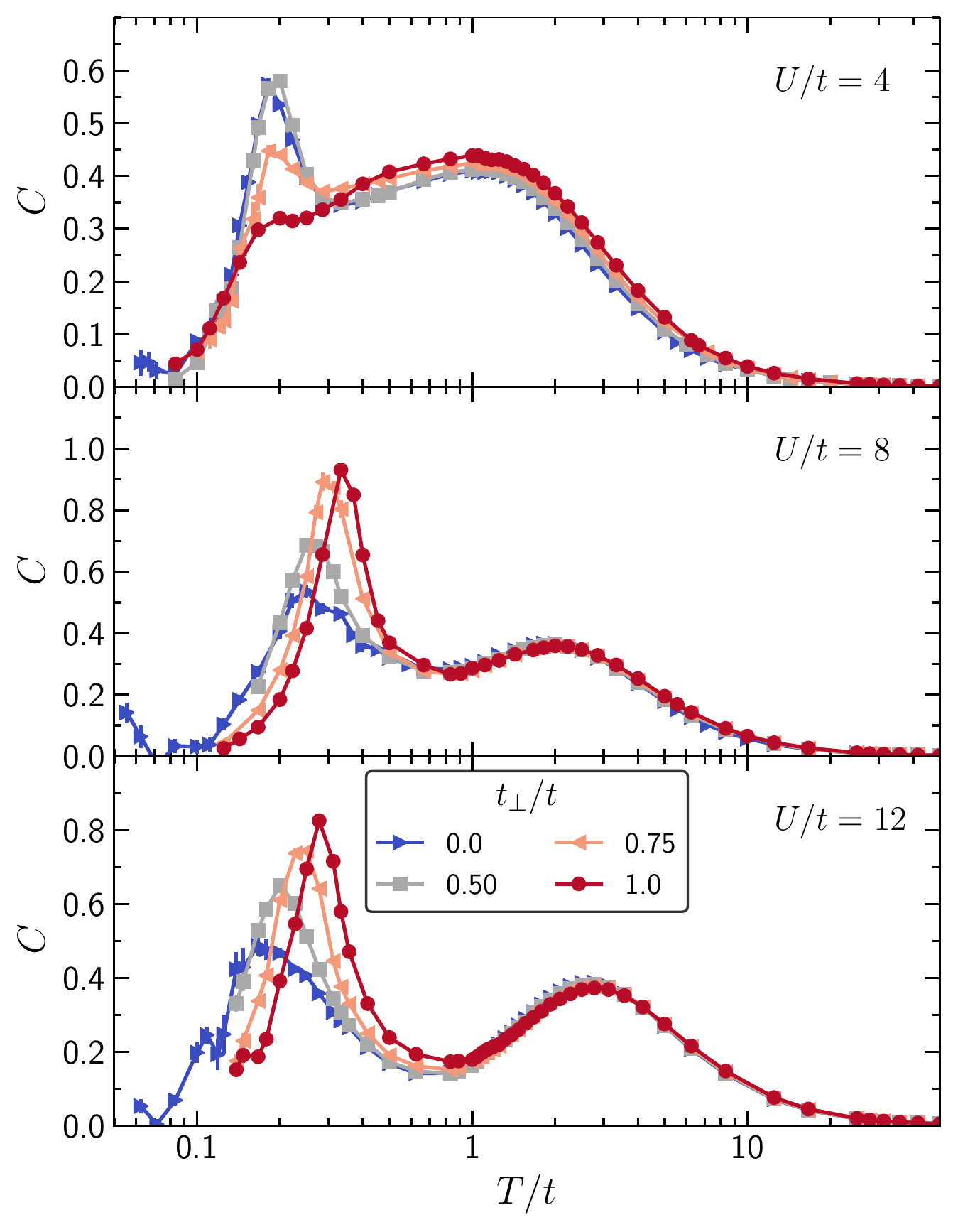}
	\caption{The specific heat $C$ for ${U/t=4,8,12}$ at different values of $t_\perp/t$. For $U/t=8,12$ the low temperature peak associated with spin degrees of freedom is reduced by anisotropy, while $U/t=4$ shows the opposite effect.  \label{fig:CT}}
\end{figure}

The specific heat as a function of temperature is a useful thermodynamic observable, showing peaks that characterize the entropy reduction as degrees of freedom reorganize and cease to fluctuate. In particular, there is a two-peak structure, shown in Fig.~\ref{fig:CT}, where at large $U/t$ one peak is associated with the charge (i.e. density) and the other with the spin degree of freedom. It is even more informative to break its contributions into the interaction energy ($P=U\mathcal{D}$) and kinetic energy ($K$) contributions.

Reference~\cite{Paiva2001} examined the contributions $dP/dT$ and $dK/dT$ to the specific heat in the 2D Hubbard model. One reason this is useful is that the interaction energy directly captures the charge fluctuations of freedom, while the kinetic energy is closely related to the spin degree of freedom (at least at large $U/t$). For $U/t=10$, the high $T$ charge peak originated in $dP/dT$ (moment formation) and the low $T$ spin peak in $dK/dT$ was related to moment ordering.  However, although the two peak structure in $C$ was clearly evident at $U/t=2$, the high $T$ peak came from $dK/dT$ and the low $T$ peak from $dP/dT$.  (The designation of these peaks as charge and spin thus clearly becomes inappropriate as $U$ gets small.) At $U/t=10$, in addition to the high $T$ peak, $dP/dT$ also had a negative dip at lower $T$.  This has also been observed in the 1D Hubbard model~\cite{Shiba1972} and dynamical mean field studies~\cite{Georges1993}. 

\begin{figure}[tbp!]
	\includegraphics[width=\linewidth]{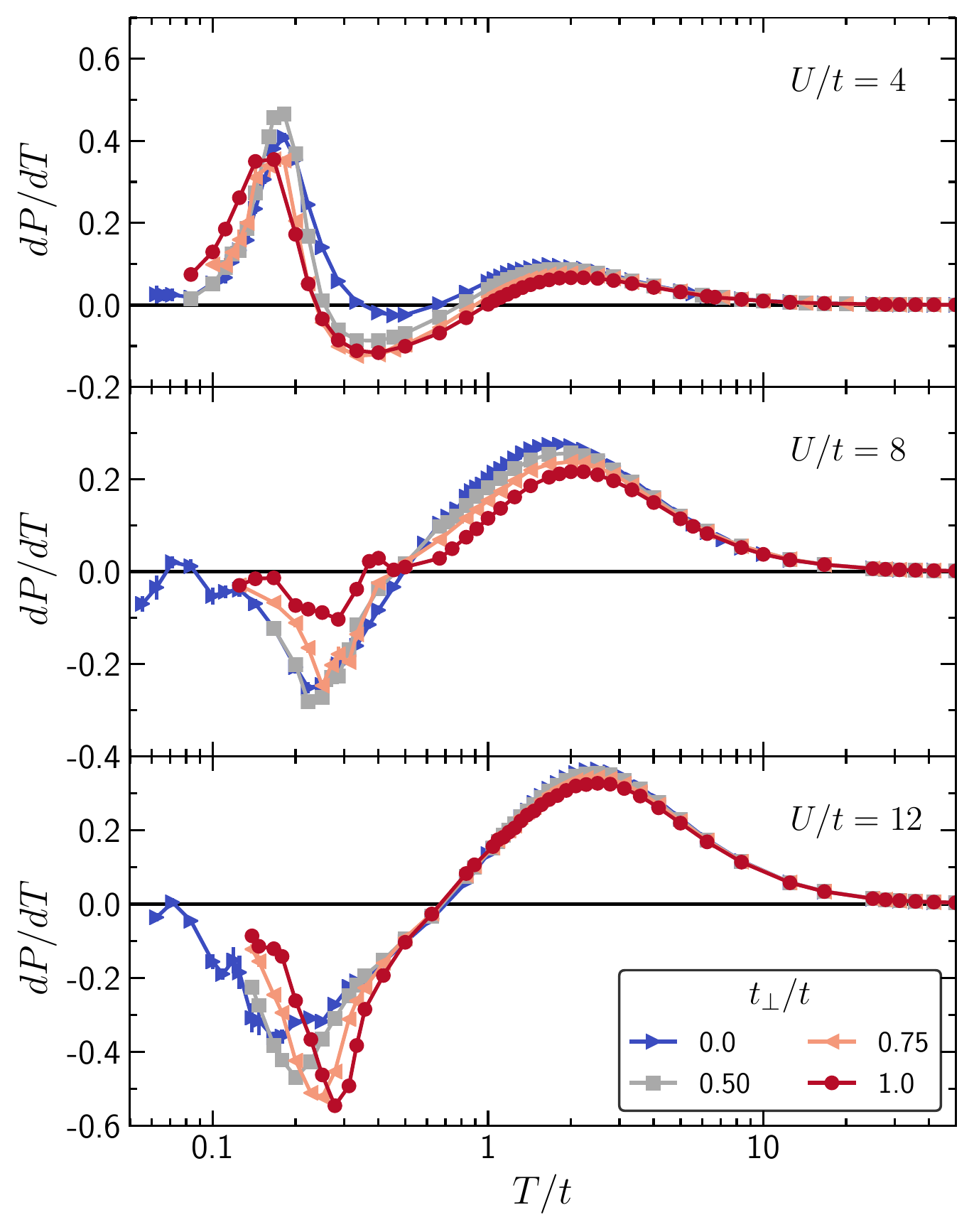}
	\caption{$dP/dT$ with $P$ the interaction energy as a function of temperature for ${U/t=4,8,12}$ at different values of $t_\perp/t$. \label{fig:dPdT}
	}
\end{figure}

\begin{figure}[tbp!]
	\includegraphics[width=\linewidth]{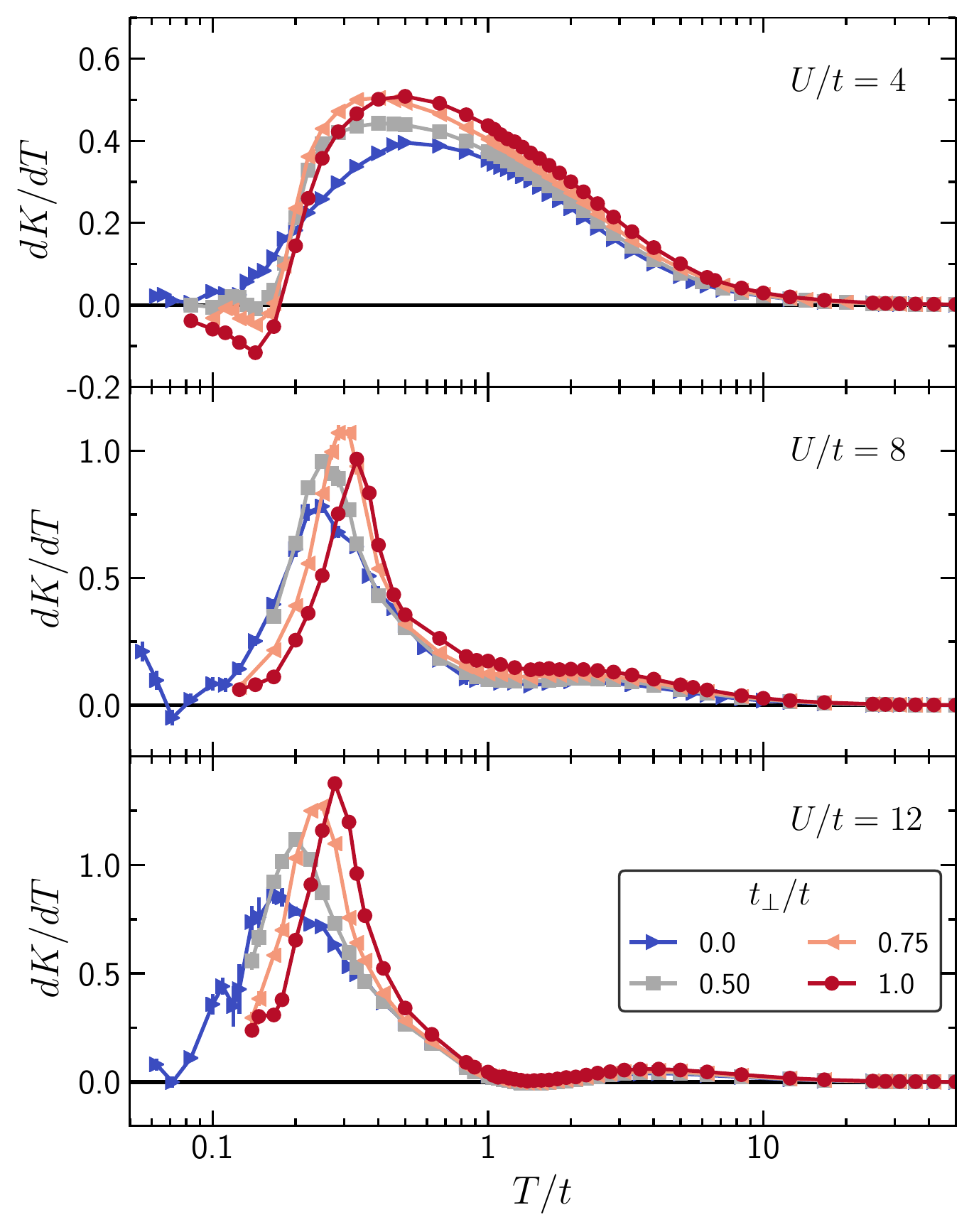}
	\caption{$dK/dT$ with $K$ the kinetic energy as a function of temperature for ${U/t=4,8,12}$ at different values of $t_\perp/t$.\label{fig:dKdT}}
\end{figure}

We show a similar decomposition of the specific heat into $dP/dT$ and $dK/dT$ in Figs.~\ref{fig:dPdT}  and~\ref{fig:dKdT} (see footnote~\footnote{In order to take derivatives in an unevenly spaced dataset, we have used three-point differentiation rule with a $\mathcal{O}(h^2)$ error, where $h$ is the spacing between the variable differentiated with respect to:
\[
f'(x) = \left[\frac{x_i - x_{i+1}}{(x_{i-1} - x_i)(x_{i-1} - x_{i+1})} \right] f(x_{i-1}) 
\] 
\[
+ \left[\frac{2x_i - x_{i-1} - x_{i+1}}{(x_i - x_{i-1})(x_i - x_{i+1})} \right] f(x_{i}) 
\]
\[
+ \left[\frac{ x_i - x_{i-1} }{(x_{i+1} - x_{i-1})(x_{i+1} - x_i)} \right] f(x_{i+1}).
\]
Error bars are obtained by error propagation and treating the errors in quadrature.} for details on the differentiation procedure). 
Figure~\ref{fig:dPdT} shows the interaction energy contribution to the specific heat, $dP/dT$. The $U/t=8$ and the $U/t=12$ data have a high temperature charge peak and a negative dip at lower $T/t$, associated with the increase in interaction energy which occurs with the formation of AF order. For $U/t=8$ the negative dip increases by more than a factor of two moving away from 3D, while for $U/t =12$ the magnitude of the dip decreases moving away from 3D, and the dip shifts to lower $T/t$ as the system becomes more 2D. Although $U/t$ is constant, $U/t_\perp$ increases as $t_\perp$ decreases; the more pronounced dip can thus be explained by an increase in the effective interaction strength. Finally, for $U/t=4$ the low temperature peak in $dP/dT$ leads to the low temperature peak in the specific heat.


The low temperature spin peak in $dK/dT$ can be seen in Fig.~\ref{fig:dKdT} for $U/t=8$  and $U/t=12$. It is mostly independent of $t_\perp/t$ although the peak position moves down in $T/t$ as the system becomes more 2D. For $U/t=4$ the peak is replaced by a broader bump that moves to higher $T/t$ as $t_\perp/t$ decreases.

Together, $dK/dT$ and $dP/dT$ combine to form the characteristic two peak structure of the specific heat seen in  Fig.~\ref{fig:CT}.  For strong couplings $U/t=8,12$ the low $T$ peak in the specific heat comes from the kinetic energy peak, and the role of the interaction energy is to reduce the height of the peak.  For $U/t=4$ we can see that both $dP/dT$ and $dK/dT$ give a positive contribution to the low $T$ peak in the specific heat. 

The interpretation of the multi-peak structure of the specific heat data is complicated by the possibility that the spin-ordering peak might itself be split owing to the presence of two distinct superexchange energy scales, $J$ and $J_\perp$.  For ${J_\perp < J}$ stochastic series expansion (SSE) studies of the 2D-3D crossover of the spin-1/2 Heisenberg model \cite{Sengupta2003} have shown the existence of a broad peak from short range 2D order, as well as a sharper 3D ordering peak whose height diminishes as $J_\perp/J$ decreases.  Resolving these structures is already challenging for the spin model, even though the SSE approach scales linearly with the number of spins $N$ and system sizes as large as $N=3 \times 10^4$ were investigated, and is not possible for the more challenging itinerant Hubbard model studied here.

\begin{figure}[tbp!]
	\includegraphics[width=0.8\linewidth]{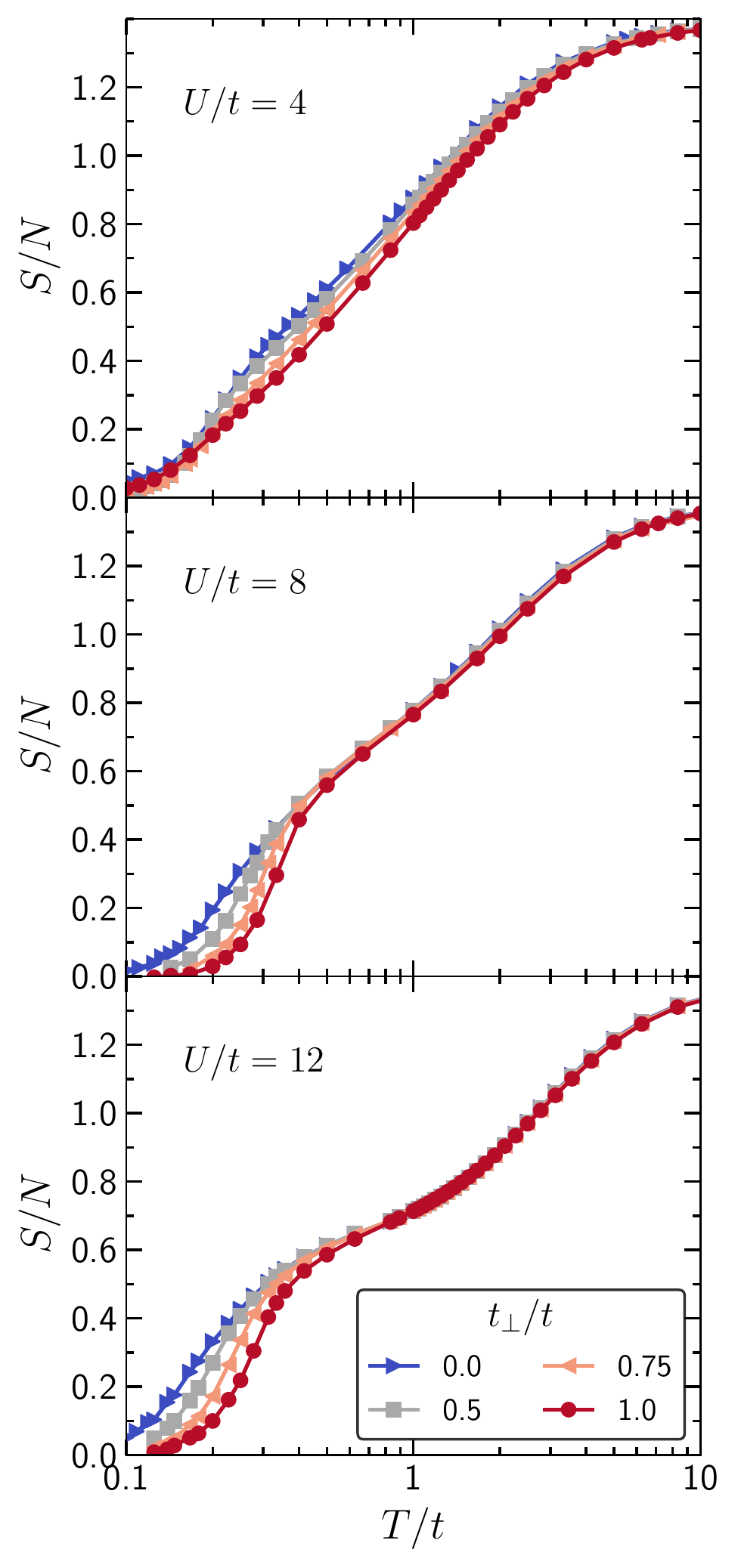}
	\caption{Entropy $S$ versus temperature for different interplane hopping $t_\perp$. Adiabatic cooling is observed as $t_\perp$ is decreased for all values of the interaction strength.\label{fig:SvsT1}}
\end{figure}

The entropy as a function of temperature has, in principle, similar information to the specific heat, but the physics is less directly apparent, as seen in Fig.~\ref{fig:SvsT1}. We compute the entropy by integrating ${dS = dQ/T = C/T \, dT}$, with ${C = dE /dT}$ the specific heat. Integrating by parts, that integral can be rewritten in terms of the energy $E$,
\begin{equation}\label{eq:S_DQMC_1}
S(T) = 2 \log(2) + \frac{E(T)}{T} - \int_T^{\infty} \frac{E(T')}{{T'}^2} dT'. 
\end{equation}
In practice, we obtain DQMC results up to a temperature cutoff ${T_{\mathrm{cut}}= 250 t}$ and use the leading order high temperature series term ($t=0$) in the integral in Eq.~\eqref{eq:S_DQMC_1} for $T>T_{\text{cut}}$ to accelerate convergence \footnote{The error in the entropy calculation due to the finite value of the temperature cutoff $T_{\mathrm{cut}}$ was estimated by comparing the results obtained with ${T_{\mathrm{cut}}/t = 100, 250}$. The difference between those two is below $3.5\times10^{-4}$ for all interaction strengths, temperatures, and values of $t_\perp$ considered in this manuscript.}.

Figure~\ref{fig:SvsT1} shows the entropy per site $S/N$ versus $T/t$ for different $t_\perp/t$ at $U/t=4,8,12$. Systems with small $t_\perp/t$ have larger $S$ for a given $T/t$.  For $U/t=4$, $S(T)$ for different values of $t_\perp/t$ begins to become distinct at ${T/t \lesssim 5}$, and then again become independent of $t_\perp$ at ${T/t \lesssim 0.1}$.  For $U/t=8,12$, the dependence on $t_\perp/t$ is negligible until ${T/t \lesssim 0.5}$. Decreasing $t_\perp/t$ at fixed entropy lowers the temperature.

\begin{figure}[tbp!]
	\includegraphics[width=\linewidth]{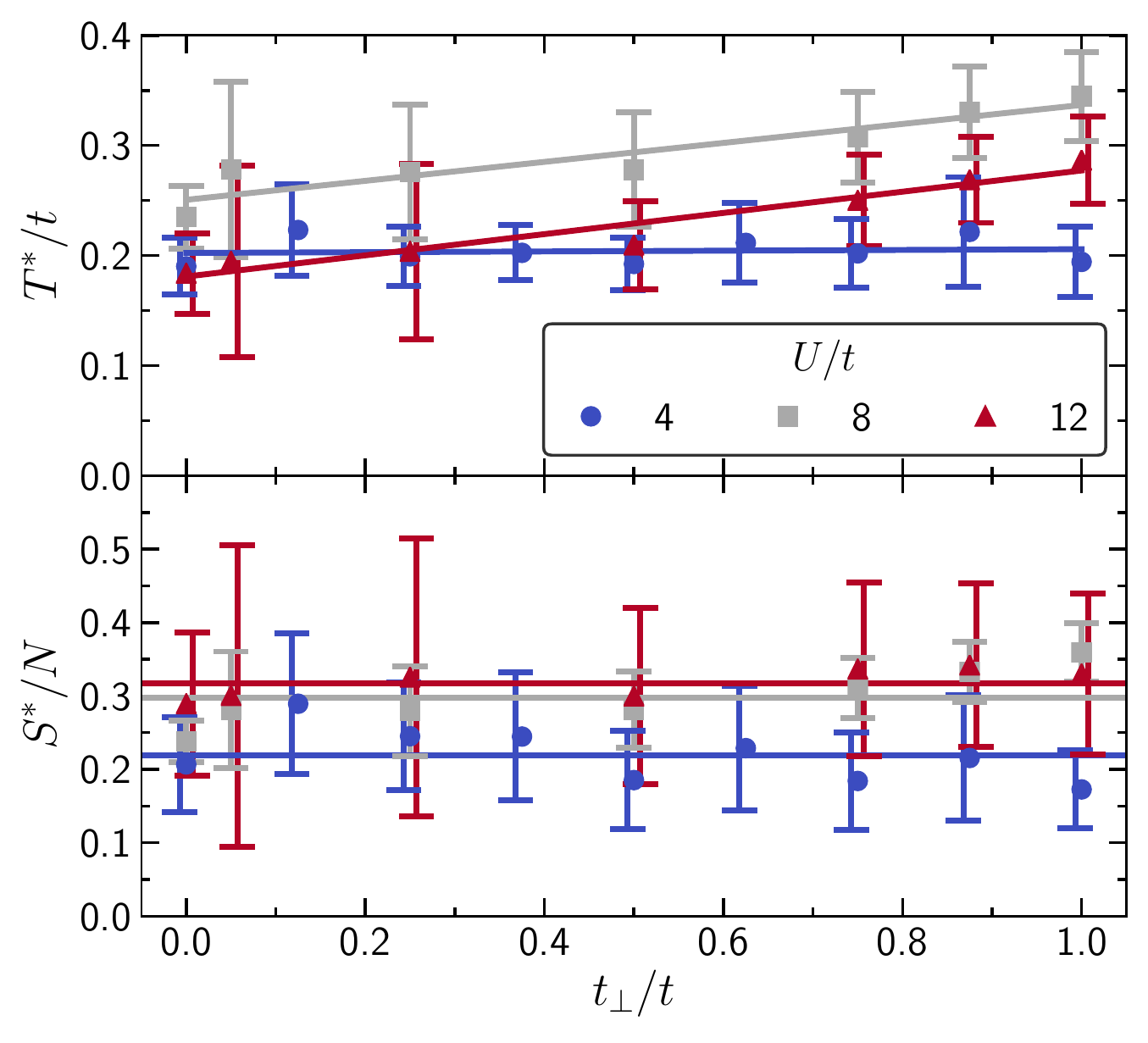}
	\caption{Top panel: $T^*$, defined as the position of the small-temperature peak in $C$; lower panel $S^*=S(T^*)$ as a function of $t_\perp/t$. Lines in the upper panel are linear fits to the data, while lines in the lower panel are the average of the datasets. \label{fig:TS_star}
	}
\end{figure}

We define the temperature for the low-$T$ peak in  $C(T)$  as $T^*$. For $t_\perp/t=1$ and $U/t=8$, $T^*$ closely coincides with the N{\'e}el temperature; while for $U/t=4$, $T^*$ is nearly in agreement with the upper bound given by Ref.~\cite{Kozik2013}. For $U/t=12$, we do not know of literature where finite-size scaling is done to extract $T_{\rm N\acute{e}el}$. For the 2D system, $T_{\rm N\acute{e}el}=0$ due to the Mermin-Wagner theorem, but in contrast ${T^* \neq 0}$.  It is also useful to define  $S^*=S(T^*)$.  Figure~\ref{fig:TS_star} shows $T^*$  and $S^*$ as functions of $t_\perp/t$.  For $U/t=8$ and $U/t=12$, $T^*$ increases with $t_\perp/t$, signaling that the formation of strong AF correlations moves to lower $T$ as $t_\perp/t$ is decreased. For $U/t=4$, on the
other hand, $T^*$  is almost independent of $t_\perp/t$. 

In the strong-coupling (Heisenberg) limit of the 2D-3D crossover, $T_{\rm N\acute{e}el}/J$ is known \cite{Chakravarty1988} to go as ${T_{\rm N\acute{e}el} \sim -1/{\rm ln}\,\alpha}$ for $\alpha \ll 1$, with $\alpha=J_\perp/J$. The isotropic case, $\alpha=1$, has the highest transition temperature ${T_{\rm N\acute{e}el}/J \sim 0.946}$~\cite{Sandvik1998}. $T_{\rm N\acute{e}el}$ decreases slowly with $\alpha$ over most the range from $0$ to $1$, then rapidly drops to zero as $\alpha\to 0$. Similar trends are observed for $T^*$ in Fig.~\ref{fig:TS_star} for large $U/t$, where the results indicate that although $T^*$ is on the same order as the 3D value at weak $t_\perp/t$, it still reaches is largest value at the isotropic point ${t_\perp=t}$. On the other hand, this strong coupling behavior does not extend to weaker coupling, as the ${U/t=4}$ data demonstrate in Fig.~\ref{fig:TS_star}, where $T^*$ is nearly independent on anisotropy. A possible explanation for this behavior is that for small $U/t$ band structure effects such as the van Hove singularity in the 2D density of states become relevant. 

\begin{figure}[tbp!]
	\includegraphics[width=\linewidth]{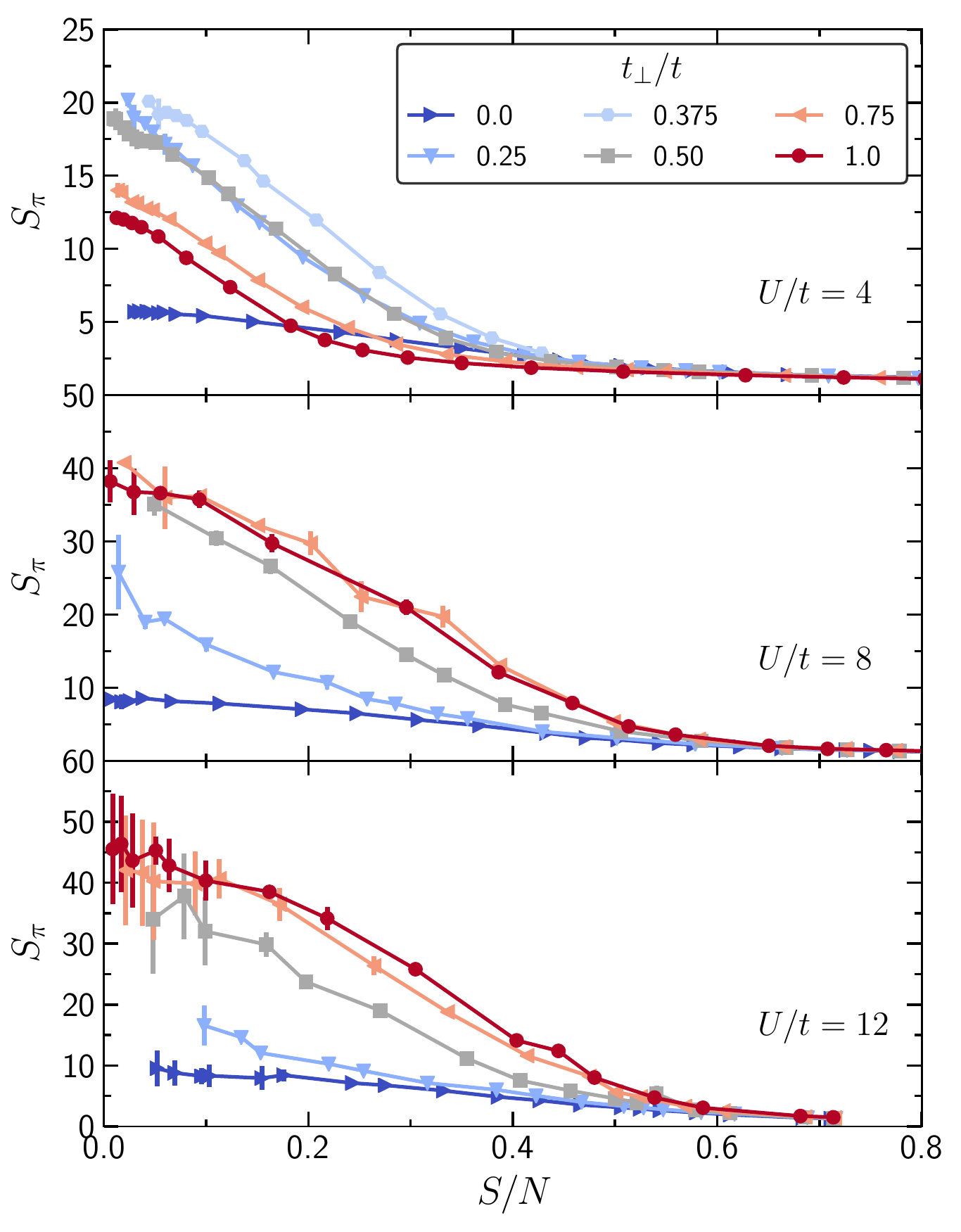}
	\caption{AF structure factor $S_\pi$ as a function of entropy $S$ for
		different interplane hopping $t_\perp/t$.  At weak coupling $S_\pi$
		grows as one moves adiabatically away from the isotropic 3D limit down
		to $t_\perp/t=0.375$ and then comes back down. \label{fig:SpipipivsS}}
\end{figure}

Previous work \cite{Gorelik2012} examined {\it short-range} magnetic order in different dimensions and concluded that for strong couplings their onset occurs at a common (dimension independent) entropy, roughly ${S/N \sim \ln 2}$. This result is in agreement with Fig.~\ref{fig:SpipipivsS} for $U/t=8,12$, where the onset of growth of the structure factor begins around ${S/N \sim \ln 2}$. That trend, however, does not extend to smaller $U/t$, as the $U/t=4$ panel  in Fig.~\ref{fig:SpipipivsS} shows.

The reduction in $S_\pi$ with anisotropy at ${U/t=8, 12}$ overwhelms the benefits of adiabatic cooling, as seen in Fig.~\ref{fig:SpipipivsS}.  At fixed entropy, $S_\pi$ is reduced by anisotropy.  In contrast, at $U/t=4$, Fig.~\ref{fig:SpipipivsS}, $S_\pi$ can be enhanced by more than a factor of two by reducing $t_\perp/t$ away from the 3D limit. As discussed previously, however, the value is never as large as the maximum attained for $U/t=8$ in the isotropic case at the same entropy.

\section{Conclusions}\label{sec:Conclusions}

We have evaluated the entropy dependence of the AF structure factor $S_\pi$ of the half-filled repulsive Hubbard model in the 2D-3D crossover, tuned by an interplanar tunneling $t_\perp$ which is less than the intraplane $t$. At interaction ${U/t=4}$ and ${T/t \lesssim 0.2}$, $S_\pi$ is maximized at intermediate ${t_\perp/t \sim 0.4}$. At stronger coupling, ${U/t = 12}$, $S_\pi$ is largest at the isotropic 3D point, $t_\perp/t = 1$; while for ${U/t=8}$,  $S_\pi$ exhibits a plateau between ${0.5 \leq t_\perp/t \leq 1}$. 

Although anisotropy enhances magnetism at $U/t =4$, the structure factor is smaller than it is for larger $U/t$ at the isotropic point ${t_\perp=t}$. Furthermore, despite some adiabatic cooling when reducing $t_\perp$ for large $U/t$, $S_\pi$ remains roughly the same for $t_\perp/t \in (0.5,1.0)$ for $U/t=8$, and diminishes with anisotropy for $U/t=12$, so there is no benefit in using anisotropy.

The study of anisotropy in the tunneling of the Hubbard model, and its strong-coupling Heisenberg limit, is of interest beyond OLE.  QMC simulations of bilayer Hubbard \cite{Scalettar1994} and Heisenberg \cite{Wang2006} models in which $t_\perp \neq t$ or $J_\perp \neq J$ have explored quantum phase transitions between AF and singlet phases relevant to heavy fermion magnetism, as well as studied $s\pm$-wave superconductivity \cite{Bang2008,Korshunov2008,Wang2009}. Similarly, the possibility of enhanced transition temperatures to magnetic order at the 2D surface of bulk 3D materials has been investigated \cite{Falicov1990,Fadley1992}. Finally, analogous issues concerning the effect of inhomogeneous intersite tunneling occur in the context of optimizing $d$-wave pairing in the 2D Hubbard Hamiltonian. In that case, a model of $2\times2$ plaquettes \cite{Scalapino1996} with internal hopping $t$ and coupled by interplaquette hopping $t'$ was suggested to have an optimal tunneling for pairing which occurs at ${t'<t}$, away from the isotropic limit \cite{Arrigoni2004,Martin2005,Kivelson2007,Tsai2006,Roising2018,Wachtel2017,Baruch2010}. 
The interest in anisotropic tunnelings also extends to the attractive Hubbard model as well. For example, in Ref.~\cite{Wachtel2012} a layer of disconnected attractive Hubbard sites coupled to a metallic layer shows that although the superconducting critical temperature $T_c$ exhibits a maximum as function of the interlayer tunneling, the highest $T_c$ is still smaller than the maximal $T_c$ of the uniform 2D attractive Hubbard model. The results presented in the present paper provide additional information in this broader context, both by quantifying how AF evolves for layered materials, and also by providing further insight into how the strong correlation physics interplays with anisotropy.

Finally, a possible application of our results is to design a cooling protocol, relying on the results of Fig.~\ref{fig:SvsT1} that show a system at a fixed entropy will get colder as $t_\perp/t$ is reduced, specially for strong interactions. By exploiting inhomogeneity, this effect can be used to cool systems with an arbitrary $t_\perp/t$, even isotropic 3D systems, as follows. First, load the atoms into a 3D lattice. Now adjust the lattice depth of the system in a carefully constructed inhomogeneous way; for simplicity think of two regions: $R$, an entropy reservoir we will sacrifice to cool the system, and $S$, the system we want to cool and study. In $R$, we adiabatically lower the $z$-direction lattice depth $V_z$. This spatially inhomogeneous lattice depth could be engineered using, for example, a spatial light modulator (however, implementing the spatially-modulated anisotropy will be more challenging than a spatially-modulated trapping potential). The now-anisotropic $R$ can carry extra entropy at a given temperature, as per Fig.~\ref{fig:SvsT1}, so entropy will transport to this region from $S$ as the system reaches thermal equilibrium at a new temperature. At the temperatures plotted for $U/t=12$, the entropy per particle in region $S$ can be reduced by a factor of 2. Finally, one can cool and study $S$ with an arbitrary $t_\perp/t$ this way by applying an optical barrier to turn transport off between $S$ and $R$, and then adiabatically change $V_z$ in the $S$ region to give the desired $t_\perp/t$. This cooling method bears similarities to other entropy redistribution protocols \cite{Ho2009,Ho2009a,Bernier2009,McKay2011,Haldar2014,Schachenmayer2015,Goto2017,Kantian2018,Chiu2018,
Mazurenko2017,Mirasola2018,VenegasGomez2020,Werner2019,Yang2020} but overcomes some difficulties. In particular, schemes that rely on metal reservoirs created by changing the local potential, rather than lattice anisotropy, suffer at large $U/t$ from the fact that the metals created this way are bad metals, therefore they carry significantly less entropy, than, e.g., a non-interacting metal. Our protocol also has some similarities to the conformal cooling suggested in Ref.~\cite{Zaletel2016}, but allows one to cool the full Fermi-Hubbard model in a practical way, rather than just the Heisenberg limit. 

\begin{acknowledgments}
The work of K.R.A.H. and E.I.G.P. was supported in part by the Welch Foundation through Grant No.~C-1872 and the National Science Foundation through Grant No.~PHY1848304. The work of R.T.S. was supported by the grant DE‐SC0014671 funded by the U.S. Department of Energy, Office of Science. The  work of R.G.H. was  partially  supported  by  the  NSF  (Grant  No.  PHY-1707992),  the  Army  Research  Office  Multidisciplinary  University Research  Initiative  (Grant  No.  W911NF-14-1-0003),  the  Office  of Naval Research, and The Welch Foundation (Grant No. C-1133). T.P. thanks the CNPq, the FAPERJ and the INCT on Quantum Information. R.M. acknowledges the funding received from the QuantERA ERA-NET Cofund in Quantum Technologies implemented within the European Unions Horizon 2020 Programme and from EPSRC under the grant EP/R044082/1.
\end{acknowledgments}

\appendix*
\section{Finite-size errors}\label{App:Appendix_A}


As mentioned in Section~\ref{sec:Hubbard_DQMC}, for the energy, kinetic energy, interaction energy (number of doublons), nearest neighbor spin correlations, and  entropy, finite size errors (as measured by the difference between $L=4$ and $L=6$ calculations) are $\lesssim 5\%$. It is the correlations at distances comparable to the system size that are affected; other than these, $S_\pi$ is thus the only observable that is affected, and only when the system is near or below the N\'eel temperature so that the correlations at separations comparable to the system size are appreciable.

In order to give an estimate of the finite-size effects for the different values of $t_\perp/t$ and $U/t$, we present $S_\pi^{(L)}$ and ${\tilde{S}_\pi^{(L)}=S_\pi^{(L)}/L^3}$ as a function of $U/t$, $t_\perp/t$ and $T/t$ in $L\times L \times L$ cubic systems. We note that for $T>T_{\text{N{\'e}el}}$ we have $S_\pi \to \text{constant}\ne  0$ as $L\to \infty$ and ${\tilde S}_\pi \to 0$, while for  $T<T_{\text{N{\'e}el}}$ we have $S_\pi \to \infty$ and ${\tilde S}_\pi \to \text{const}$.

\begin{figure}[tbp!]
	\includegraphics[width=\linewidth]{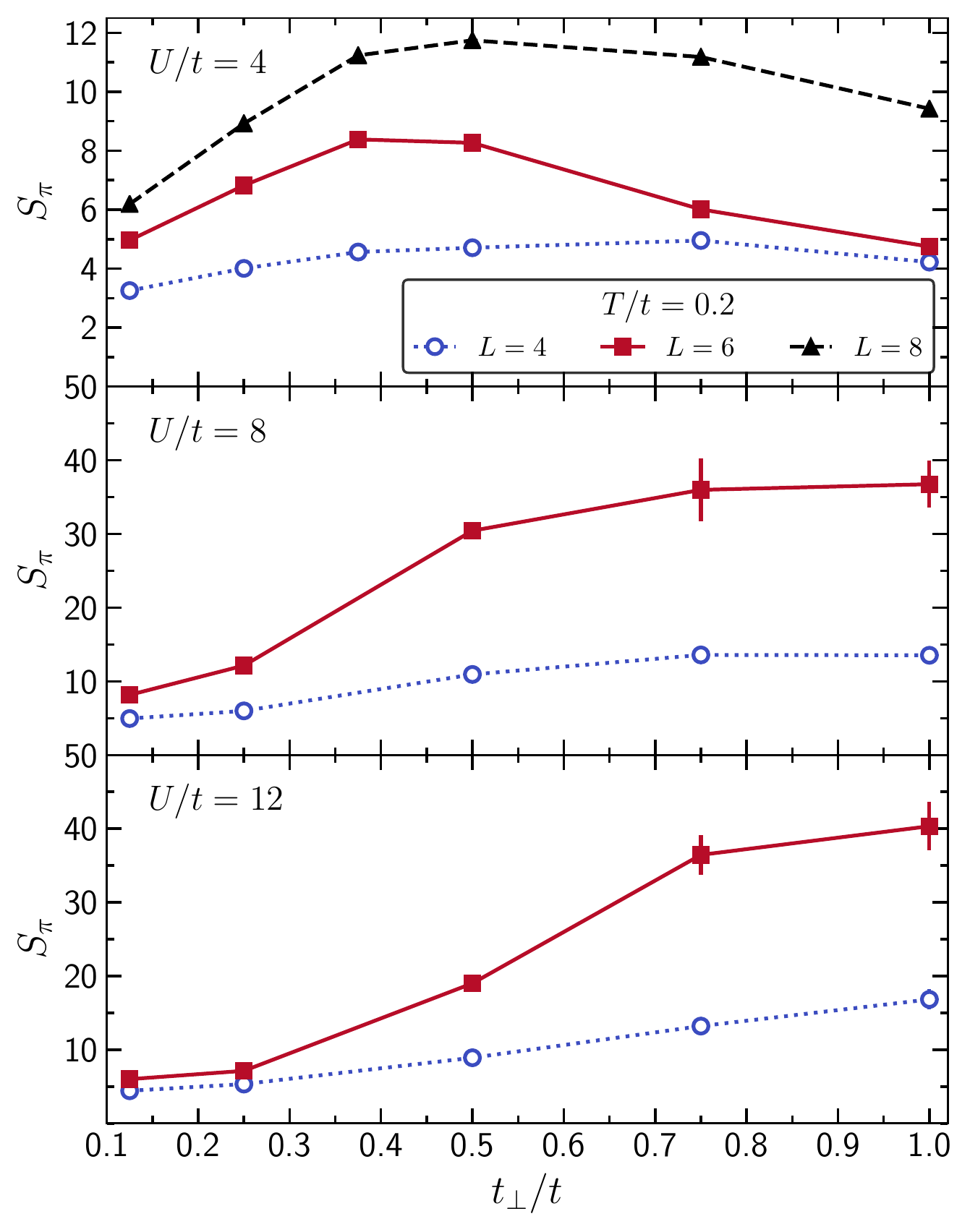}
	\caption{AF structure factor $S_\pi$ as a function of interplane tunneling 
	    $t_\perp/t$ at $T/t=0.2$ for different cubic lattices with sides of length $L$. We note that the $L=8$ calculations are substantially more computationally expensive because -- defining $N_{\text{sites}}=L^3$ to be the number of sites -- the computational cost scales as $O(N_{\text{sites}}^3)=O(L^9)$.   \label{fig:Spivstz_FSS}}
\end{figure}

Fig.~\ref{fig:Spivstz_FSS} presents $S_\pi$ at ${T/t=0.2}$ as a function of $t_\perp/t$ for different system sizes, while Tables~\ref{table:fss} and \ref{table:fss2} report ${\tilde S}_\pi$ for two $T/t$ presented in Fig.~\ref{fig:Spivstz}. The ${U/t=8,12}$ panels exhibit the same behavior seen in Fig.~\ref{fig:Spivstz}, i.e. $S_\pi$ is maximized when ${t_\perp \sim t}$. For $t_\perp/t \gtrsim 0.5$, $S_\pi$ grows roughly proportional to $L^3$, suggesting that the system is below $T_{\text{N{\'e}el}}$, and that the numerics provides a reasonable estimate of ${\tilde S}_\pi$. In contrast, the $U/t=4$ panel demonstrates that $S_\pi$ is maximized at the 2D-3D crossover, in agreement with the results presented in Fig.~\ref{fig:Spivstz} although the location of the maximum depends significantly on the system size. The scaling looks neither like a the simple $L$-independent $S_{\pi}$ expected in large systems for temperature above the N{\'e}el temperature, nor the $L$-independent ${\tilde S}_{\pi}$ expected for large systems below the N{\'e}el temperature. Previous results in the $t_\perp/t \ll 1$ limit \cite{Dare1996}, and in 3D \cite{Kozik2013}, place $T_{\text{N{\'e}el}} \lesssim 0.2$ for $U/t=4$; therefore this absence of a simple scaling on $L$ is expected at $T/t=0.2$. A detailed study of finite-size effects, as was done in \cite{Kozik2013,Staudt2000} in 3D, and for larger system sizes than in the present paper is required to precisely determine $T_{\rm N\acute{e}el}$ in the 2D-3D crossover. This task is out of the scope of the paper, but our results will provide a useful starting point for such calculations.



\begin{table}[tbp!] 
 \caption{Structure factor $\tilde{S}_\pi^{(L)}$ in cubic lattices with sides of length $L$ at $T/t =0.125$. \label{table:fss}}
 \begin{ruledtabular}
 \begin{tabular}{l l l l l }
 $U/t$ & $t_\perp/t$ & $\tilde{S}_\pi^{(4)}$ & $\tilde{S}_\pi^{(6)}$ & $|\tilde{S}_\pi^{(6)} - \tilde{S}_\pi^{(4)}|$ \\
 \hline
 \hline
\multirow{6}{*}{4} & 0.125 & 0.06 & 0.04  &  0.03 \\ 
							 & 0.25   & 0.10 & 0.08  & 0.02 \\
							 & 0.375 & 0.11 & 0.09  & 0.02 \\
 							 & 0.5     & 0.11 & 0.08  & 0.03\\
 							 & 0.75   & 0.11 & 0.06  & 0.05 \\
 							 & 1        & 0.09 & 0.05  & 0.04 \\ 
 \hline
 \multirow{5}{*}{8} & 0.125 & 0.09  & 0.05 & 0.04 \\
 							 & 0.25    & 0.12  & 0.09 & 0.03 \\ 
                            & 0.5      & 0.20  & 0.17 & 0.03 \\
 							 & 0.75    & 0.22  & 0.18 & 0.04 \\
 							 & 1         & 0.21  & 0.17 & 0.04 \\ 
 \hline
 \multirow{5}{*}{12} &  0.125 & 0.09 & 0.05 & 0.04 \\
                               &  0.25   & 0.11 & 0.08 & 0.03 \\
 								&  0.5     & 0.20 & 0.16 & 0.04 \\
 								&  0.75   & 0.26 & 0.19 & 0.06 \\
 								&  1        & 0.25 & 0.21 & 0.04 \\
 \end{tabular}
 \end{ruledtabular}
 \end{table}

\pagebreak
 \begin{table}[hbtp!] 
 \caption{Structure factor $\tilde{S}_\pi^{(L)}$ in cubic lattices with sides of length $L$ at $T/t =0.2$. \label{table:fss2}}
 \begin{ruledtabular}
 \begin{tabular}{l l l l l l l}
 $U/t$ & $t_\perp/t$ & $\tilde{S}_\pi^{(4)}$ & $\tilde{S}_\pi^{(6)}$ & $\tilde{S}_\pi^{(8)}$ & $|\tilde{S}_\pi^{(6)} - \tilde{S}_\pi^{(4)}|$ & $|\tilde{S}_\pi^{(8)} - \tilde{S}_\pi^{(6)}|$ \\
 \hline
 \hline
\multirow{6}{*}{4}  & 0.125 & 0.05 & 0.02 & 0.01 & 0.03 & 0.01 \\ 
							  & 0.25   & 0.06 & 0.03 & 0.02 & 0.03 & 0.01 \\
							  & 0.375 & 0.07 & 0.04 & 0.02 & 0.03 & 0.02 \\
							  & 0.5     & 0.07 & 0.04 & 0.02 & 0.04 & 0.02 \\
							  & 0.75   & 0.08 & 0.03 & 0.02 & 0.05 & 0.01 \\
							  & 1        & 0.07 & 0.02 & 0.02 & 0.04 & 0.00  \\ 
\hline
 \multirow{5}{*}{8} &  0.125 & 0.08 & 0.04 && 0.04 & \\
 							    & 0.25 & 0.09 & 0.06 && 0.04 &\\
								& 0.5 & 0.17 & 0.14 && 0.03 &\\
 								&  0.75 & 0.21 & 0.17 && 0.05 &\\
 								&  1 & 0.21 & 0.17 && 0.04 & \\
 \hline
 \multirow{5}{*}{12} & 0.125 & 0.07 & 0.03 && 0.04 & \\
 							    & 0.25 & 0.08 & 0.03 && 0.05 & \\ 
 								& 0.5 & 0.14 & 0.09 && 0.05 & \\
 								& 0.75 & 0.21 & 0.17 &&  0.04 &\\
 								& 1 & 0.26 & 0.19 && 0.08  & \\
 \end{tabular}
 \end{ruledtabular}
 \end{table}

\newpage
\bibliography{2D3Dcrossover.bib}

\end{document}